\documentclass{aa}
\usepackage[varg]{txfonts}
\bibpunct{(}{)}{;}{a}{}{,} 
\usepackage{graphicx}	
\usepackage{amsmath}	
\usepackage{natbib}

\begin{document}

\title{\texttt{RTModel}: a platform for real-time modeling and massive analysis of microlensing events}

\author{V. Bozza\inst{1,2}}

\offprints{V. Bozza, \email{valboz@sa.infn.it}}

\institute{Dipartimento di Fisica "E.R. Caianiello", Universit$\grave{a}$ di Salerno, Via Giovanni Paolo  132, Fisciano, I-84084, Italy 
\and Istituto Nazionale di Fisica Nucleare, Sezione di Napoli, Via Cintia, Napoli, I-80126, Italy}

\date{Received/ Accepted}

\abstract {Microlensing of stars in our Galaxy has long been used to detect and characterize stellar populations, exoplanets, brown dwarfs, stellar remnants and whatever objects may magnify the source stars with their gravitational fields. The interpretation of microlensing light curves is relatively simple for single lenses and single sources but becomes more and more complicated if we add more objects and take their relative motion into account.}{ \texttt{RTModel} is a modeling platform that has been very active in the real-time investigation of microlensing events, providing preliminary models that have proven very useful for driving follow-up resources towards the most interesting events. The success of \texttt{RTModel} is due to the ability to make a thorough and aimed exploration of the parameter space in a relatively short time.}{This is obtained by three key ideas: the initial conditions are chosen from a template library including all possible caustic crossing and approaches; the fits are performed by the Levenberg-Marquardt algorithm using a bumper mechanism to explore multiple minima; the basic computations of microlensing magnification are performed by the fast and robust \texttt{VBBinaryLensing} package.}{ In this paper we will illustrate all algorithms in \texttt{RTModel} in detail, with the purpose of fostering new ideas in view of future microlensing pipelines aimed at massive microlensing analysis.}
{}
\keywords{gravitational lensing: micro; Methods: data analysis; binaries: general; planetary systems}
\maketitle 
  
\section{Introduction}\label{intro}

Microlensing is a well-established variant of gravitational lensing in which the telescope resolution is insufficient to distinguish multiple images of a source or their distortion \citep{Mao2012}. The primary observable is the flux variation produced by the magnification of the source by the gravitational field of the lens moving across the line of sight. It was first suggested as a means to count dark compact objects contributing to the budget of the Galactic dark matter \citep{Paczynski1986}. However, it was soon realized that microlensing is particularly sensitive to the multiplicity of the lensing objects, thanks to the existence of caustics, i.e. closed curves in the source plane bounding regions in which a source is mapped in an additional pair of images \citep{Schneider1986,Erdl1993,Dominik1999}. The light curves of binary microlensing events are then decorated with abrupt peaks at caustic crossings joined by typical "U"-shaped valleys \citep{Mao1991,Alcock1999}. In addition, every time the source approaches a cusp point in a caustic, the light curve features an additional smooth peak. Such ensembles of peaks may sometimes conspire to build formidable puzzles for the microlensing modelers who aim at reconstructing the geometry and the masses of the lenses \citep{Quintuple2018}.

The emergence of microlensing as a new and efficient method to detect extrasolar planets invisible to other methods has motivated continuous observation campaigns towards the Galactic bulge that have now collected more than twenty-years of data \citep{Gaudi2012,Tsapras2018,Mroz2023}. These surveys are accompanied by follow-up observations of the most interesting microlensing events with higher cadence with the purpose to catch all elusive details of the ``anomalies'' produced by low-mass planets around the lenses \citep{Beaulieu2006}. Given the transient nature of microlensing events, it is necessary to have the maximal coverage of the light curves avoiding any gaps that might make the interpretation of the data ambiguous \citep{Yee2018}. 

Driving limited follow-up resources toward interesting events requires a real-time modeling capacity that is not trivial in a phenomenon where multiple interpretations are possible and even the computation of one light curve is time-consuming. Several expert modelers have provided many contributions over the years with their specific codes designed for efficient searches in the parameter space with the available computational resources \citep{BennettRhie1996,Bennett2010,Han2024}. Some of these codes have also been adapted for massive analysis of large microlensing datasets \citep{Koshimoto2023,Sumi2023}. 

Among the modeling platforms contributing to a prompt classification of ongoing anomalies, \texttt{RTModel} stands out with a considerable number of contributions \citep{Tsapras2014,Rattenbury2015,Bozza2016,Henderson2016,Hundertmark2018,Bachelet2019,Dominik2019,Street2019,Tsapras2019,Fukui2019,Rota2021,Herald2022}. \texttt{RTModel} was originally designed as an extension of the ARTEMiS\footnote{\url{http://www.artemis-uk.org}} platform \citep{ARTEMIS2007,ARTEMIS2008}, which provides real-time single-lens modeling of microlensing events. The original goal was to find binary and planetary models for ongoing anomalies in real-time without any human intervention. The results of the online modeling are still automatically uploaded to a website\footnote{\url{http://www.fisica.unisa.it/GravitationAstrophysics/RTModel.htm}} where they are publicly visible to the whole community. Besides real-time modeling, \texttt{RTModel} has also been used to search for preliminary models in a number of microlensing investigations \citep{Bozza2012,Bozza2016,Rota2021,Herald2022} and in the search for planetary signals in the MOA retrospective analysis \citep{Koshimoto2023}. Finally, it participated in the WFIRST microlensing data challenge\footnote{\url{https://roman.ipac.caltech.edu/docs/street_data_challenge1_results.pdf}} (see Section \ref{Sec DataChallenge}) demonstrating its potential for massive data analysis.

After so many years of service, we believe that the time has come for an article fully dedicated to \texttt{RTModel} illustrating the algorithms that contributed to its success. Inspired by some previous studies, these algorithms contain many innovative ideas tailored on the microlensing problem. We should not forget that the code for the microlensing computation on which \texttt{RTModel} is based is already publicly available as a separate appreciated package with the name \texttt{VBBinaryLensing} \citep{Bozza1,Bozza2,Bozza3}. This package was created within the \texttt{RTModel} project and is now included in several microlensing modeling platforms \citep{pyLIMA2017,MulensModel2019,muLAN2018}. In addition to that, the whole \texttt{RTModel} code has been completely revised and cleaned up and has been made public on a dedicated repository\footnote{\url{https://github.com/valboz/RTModel}}. This paper will provide a detailed description of the algorithms behind this software as a reference for users and possible inspiration of further ideas and upgrades in view of future massive microlensing surveys. 

The paper is structured as follows: Section 2 explains the global architecture of \texttt{RTModel}, with a presentation of the different modules. Section 3 deals with the pre-processing of the datasets. Section 4 illustrates the choice of initial conditions for fitting. Section 5 discusses the fitting algorithm. Section 6 shows the selection of models and the removal of duplicates. Section 7 describes the final classification of the microlensing event and the assignment to a specific class. Section 8 discusses the success rate reached in the WFIRST data challenge as a specific example demonstrating the efficiency of \texttt{RTModel} on a controlled sample. Section 9 contains the conclusions.

\section{General architecture of \texttt{RTModel}} \label{Sec Architecture}

\begin{figure}
    \centering
    \includegraphics[width = 0.5\textwidth]{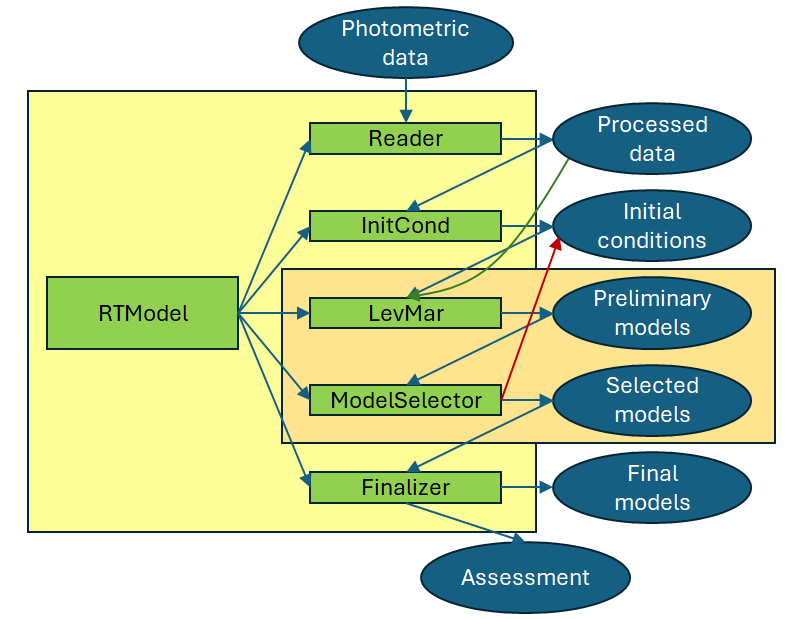}
    \caption{Flow chart of \texttt{RTModel} with the different modules in green and the data/products in blue ovals. The orange box highlights the modules that are called repeatedly for each model category with their respective products.}
    \label{fig:flowchart}
\end{figure}

\texttt{RTModel} is in the form of as a standard Python package regularly importable by Python scripts or Jupyter notebooks. 
It is made up of a master program calling a number of external modules for specific tasks. The communication between modules is ensured by human readable ASCII files. This allows an easy control of the flow and possible manual interventions if needed. Furthermore, in case of any interruptions before the conclusion of the analysis, the master program is able to automatically recover all partial results and continue the analysis up to the full completion.

While the master program is in Python, the external modules are written in \texttt{C++} language for higher efficiency. The individual modules can also be launched separately by the user, if desired. Here is the list of modules called by the \texttt{RTModel} master program, accompanied by a brief description:

\begin{itemize}
\item \texttt{Reader}: data pre-processing, including cutting unneeded baseline, re-binning, re-normalization of the error bars, outliers rejection.
\item \texttt{InitCond}: initial conditions setting, obtained by matching the peaks found in the photometry to the peaks of templates in a fixed library.
\item \texttt{LevMar}: fitting models from specific initial conditions by the Levenberg-Marquardt algorithm; multiple solutions are obtained by a bumper mechanism.
\item \texttt{ModelSelector}: Selection of best models of a given class and removal of duplicates.
\item \texttt{Finalizer}: interpretation of the microlensing event obtained by comparing the chi squares of the found models following Wilks' theorem.
\end{itemize}

Each of this module is discussed in detail in a dedicated section in the following text. Fig. \ref{fig:flowchart} shows the flow chart of \texttt{RTModel}: the master program calls the individual modules one by one. Each module takes some data as input and provides some output to be used by following modules. The two modules in the orange box are repeated for each model category (single-lens, binary-lens, ...)

\section{Data pre-processing} \label{Sec data}

The photometric series from different telescopes can be quite inhomogeneous, with different levels of scatter that might not be reflected by the reported error bars. Without any corrections, there is the danger that the fit is dominated by poor datasets with unrealistically small error bars. Furthermore, some data may still contain outliers that may alter the whole modeling process. On the other hand, some datasets may bring redundant data that add no useful information but slow down the fit process.

The \texttt{Reader} module takes care of all these aspects and tries to combine all the available photometry assigning the right weight to each data point as described below.

There should be one input file per photometric series (identified by its telescope and filter). For each data point, the input file should contain the magnitude (instrumental or calibrated), the error and the time in Heliocentric Julian Date (HJD).

Datasets are first loaded and time-ordered. It is possible to consider data from all years, to take into account multi-year baselines, or focus on one year only, if the multi-year baseline is not reliable.

\subsection{Scatter assessment} 

\begin{figure}
    \centering
    \includegraphics[width = 0.2\textwidth]{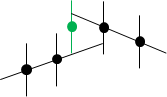}
    \caption{Assessment of scatter in the photometry. The green point is one-sigma above extrapolation from the two preceding points and half-sigma from the extrapolation of the two following points.}
    \label{fig:scatter}
\end{figure}

The scatter in each dataset is estimated by summing the residuals of each point from linear extrapolation of the two previous points and the two following points (see Fig. \ref{fig:scatter}). The residual is then down-weighted exponentially with time-distance. Let's see this in quantitative formulae.

If $f_i$ is the flux of the $i$-th point in the photometric series, $\sigma_i$ its error and $t_i$ the time of the measurement, we consider the residual from extrapolation of the previous two points as
\begin{equation}
r_i^-=\left(\frac{\Delta_i^-}{\sigma_i^-} \right)^2 w_i^-
\end{equation}
where the difference from the extrapolation is (see Fig. \ref{fig:scatter})
\begin{equation}
\Delta_i^-=f_i-\left[ f_{i-2}+c_i(f_{i-1}-f_{i-2})\right],
\end{equation}
with linear extrapolation coefficient
\begin{equation}
c_i=\frac{t_i-t_{i-2}}{t_{i-1}-t_{i-2}},
\end{equation}
and the error in the extrapolation is obtained by error propagation as
\begin{equation}
\sigma_i^-=\sqrt{\sigma_i^2+\sigma_{i-1}^2c_i^2+\sigma_{i-2}^2(1-c_i)^2}.
\end{equation}
The residual is further weighted by 
\begin{equation}
w_i^-=\exp\left[-(t_i-t_{i-2})/\tau\right],
\end{equation}
so as to avoid using information from too far points, where extrapolation makes no sense any more. The constant $\tau$ is set to 0.1 days since in most cases we do not expect features lasting less than a few hours in microlensing. Even if present (e.g. at a caustic crossing), such features are isolated and do not contribute much to the total scatter that would still be dominated by the overall behavior of the dataset.

Similar expressions hold for the residual $r_i^+$ calculated from the extrapolation of the two following points. 

The average scatter of the dataset is then
\begin{equation}
S=\sqrt{\frac{1+\sum (r_i^-+r_i^+)}{1+\sum(w_i^-+w_i^+)}}.
\end{equation}
The unity is added so as to avoid instabilities for too sparse datasets (which would come up with vanishing weights). For these, in fact, our assessment on the intrinsic scatter would be impossible.

\subsection{Outliers removal}
\begin{figure}
    \centering
    \includegraphics[width = 0.2\textwidth]{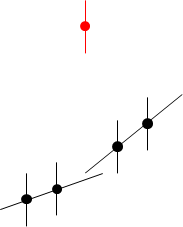}
    \caption{Identification of an outlier: if the extrapolations from the preceding and following points agree, a threshold is set for the residual of the current point. Beyond this threshold, the point is removed as outlier.}
    \label{fig:outlier}
\end{figure}

The residual from the extrapolations from previous and following points is also useful to identify and remove outliers. If the two extrapolations agree better than $3\sqrt{\left(\sigma_i^-\right)^2+\left(\sigma_i^+\right)^2}$ while the residuals exceed a user-defined threshold (default value is 10), the point is identified as outlier and removed from the dataset. In detail, the condition is $\sqrt{\left(r_i^-\right)^2+\left(r_i^+\right)^2}>thr_{outliers}$. See Fig. \ref{fig:outlier} for a pictorial view.

\subsection{Error bar re-scaling}

In the end, the error bars are rescaled by the factor $S$ tracking the average discrepancy between each point and the extrapolations from previous or following points. The error bar re-scaling proposed in \texttt{Reader} is particularly sensitive to datasets with large scatter on short time-scales. Unremoved outliers contribute to the increase of the error bars if they occur close to other data points. Scatter on longer time-scales is more difficult to identify without a noise model. At this early stage, in which we want to find preliminary models for a microlensing event, it is premature to have a more aggressive attitude. Optionally, error bar re-scaling can be turned off if all datasets are believed to have correct error bars.

\subsection{Re-binning}

\begin{figure}
    \centering
    \includegraphics[width = 0.5\textwidth]{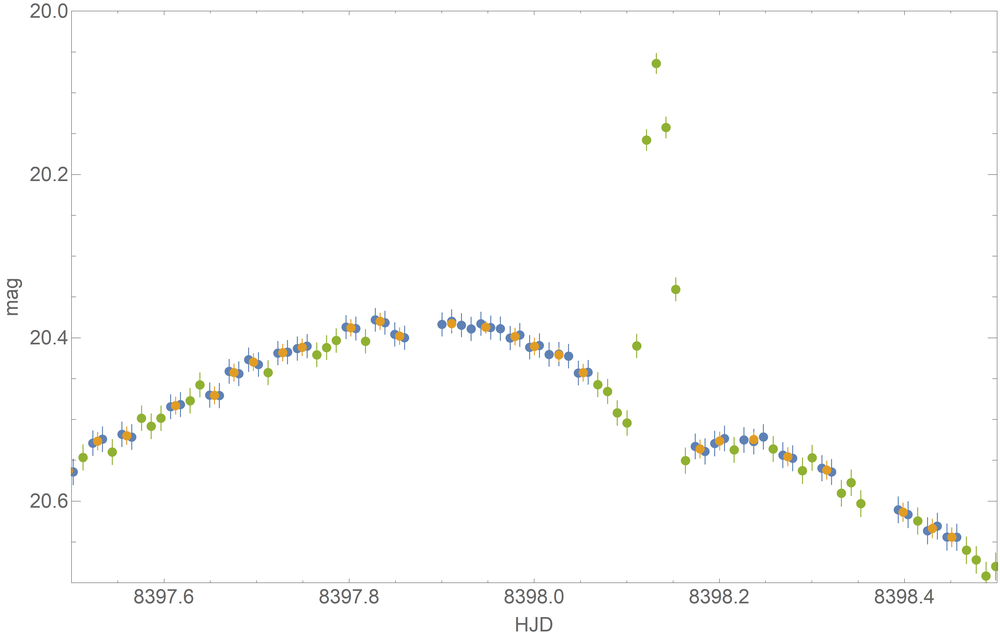}
    \caption{Re-binning at work on a simulated event. Points from the original data untouched by re-binning are in green. Original points replaced by their re-binned versions with neighbors are in blue, new points resulting from re-binning are in yellow.}
    \label{fig:re-binning}
\end{figure}

One of the goals of \texttt{RTModel} is to provide real-time assessment of ongoing microlensing events. The computational time grows linearly with the number of data points, but sometimes we have a huge number of data points on well-sampled sections of the light curve. Baseline data points also increase the computational time but their information could be easily taken into account by just few points. 

The idea pursued by \texttt{Reader} in order to speed up calculations is to replace redundant data points by their weighted mean. Data are consequently binned down to a specified number of data points. The target number can be specified by the user depending on the available computational resources and the specific problem.

The re-binning proceeds according to a significance indicator that is calculated for each pair of consecutive data points in a given dataset and estimated as
\begin{equation}
Y_i= \frac{(f_i-f_{i-1})^2}{\sigma_i^2+\sigma_{i-1}^2} + \frac{(t_i-t_{i-1})^2}{\tau^2}.
\end{equation}

In practice, a consecutive pair of data points has low significance if the two points differ less than their combined uncertainty and they are closer than the time threshold $\tau$ already mentioned before for scatter assessment. 

The pair with the lowest significance in all datasets is then replaced by a weighted mean. The significance of the new point in combination with the preceding one and the following one are re-calculated and the re-binning procedure is repeated until we are left with the desired number of points. Optionally, the significance of points that are outside the peak season (identified as the season in which the highest standard deviation is achieved) can be severely down-weighted. In this way, baseline points are strongly re-binned and computational time is saved for more interesting points.

Fig. \ref{fig:re-binning} shows an example of re-binning on a simulated event. We can see that the fundamental points describing the anomaly are preserved, while some other points where the light curve is not changing rapidly are replaced by re-binned versions along with their neighbors.

There are certainly events for which a too aggressive re-binning leads to the smoothing of tiny features in the anomaly. Typical examples are high-magnification events with many data points on the peak that are important to assess the presence of planets tracked by perturbations of the central caustic. So, re-binning must be used with moderation and keeping in mind that it has been originally introduced with the purpose of guaranteeing preliminary models in a definite short time. Nevertheless, the same philosophy of real-time modeling applies to the analysis of massive datasets as long as the purpose remains the same, i.e. a first assessment of the nature of the microlensing event. Tests on the data challenge light curves (Section \ref{Sec DataChallenge}) confirm that the computational time scales linearly with the specified number of points that are left after re-binning. Both error bar re-scaling and re-binning can be tuned to have the best performance for the specific datasets to be analyzed or can be easily turned off by the user.

The final outcome of \texttt{Reader} is a single ASCII file containing all data labeled according to their original datasets. Data taken from satellite also carry a satellite number, which will be used by the fitting module to identify them and perform the correct computation.

\section{Initial conditions setting} \label{Sec InitCond}

The strategy of \texttt{RTModel} for binary microlensing modeling is to start the fits from a finite set of initial conditions described by a template library following a philosophy initially introduced by \citet{MaoDiStefano} and pursued by \citet{Liebig} in their systematic search of light curve classes in binary microlensing. Such templates must be matched to the observed data points in order to set the initial values of the parameters for the fits. These tasks are performed by the dedicated module \texttt{InitCond}, which operates in two phases: identification of peaks in the data, template matching.

We will now discuss these two steps in this section focusing on the basic models with no higher order effects. We postpone the discussion of parallax, xallarap and orbital motion to Sec. \ref{Sec sequence}.

\subsection{Identification of peaks in the data}

Template matching is achieved by matching two peaks in the template to two analogous features in the data. So, the purpose of this stage is to find the times $t_1$ and $t_2$ of the two most prominent peaks in the data. In the absence of a second well-defined peak, a shoulder or an asymmetry in the light curve can be considered as an embryo of a peak that with a small change in the parameters would become a real peak. Therefore, the following steps are aimed at identifying such features.   


\begin{figure}
    \centering
    \includegraphics[width = 0.5\textwidth]{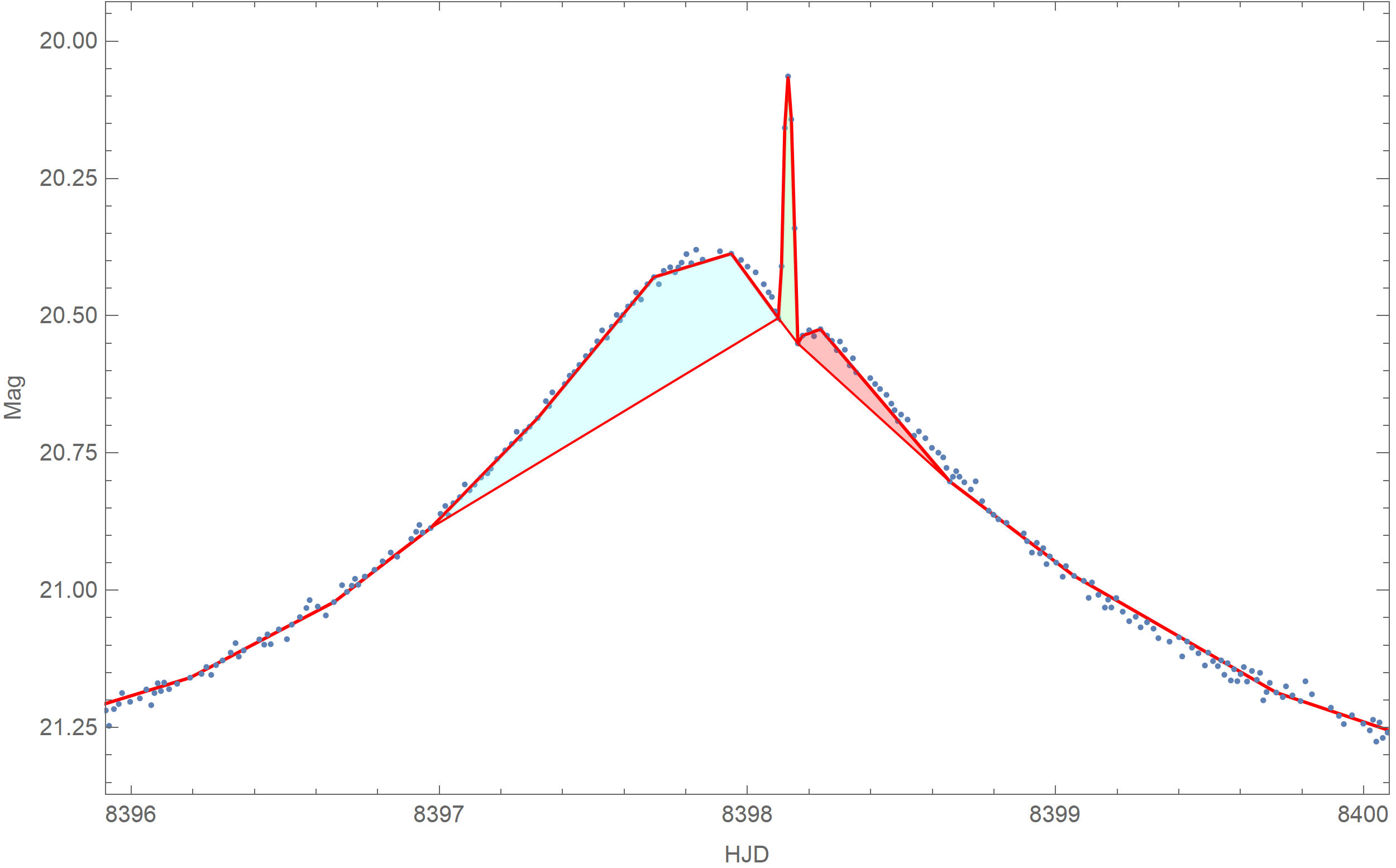}
    \caption{Spline model for a simulated dataset. Concave sections are also highlighted with different colors.}
    \label{fig:spline}
\end{figure}

\textbf{Spline modeling:} Each dataset is modeled by a linear spline in the following way. We start by a flat line at the mean flux level. Then we add the point with the highest residual from this line to the spline model. We continue adding points to the spline model until there are no points with residual higher than a chosen threshold ($5\sigma$ as default). The spline model so obtained serves as basis for the identification of peaks and shoulders in the following steps. Fig. \ref{fig:spline} shows an example of spline derived in this way from a simulated dataset.

\textbf{Concavities and convexities: } A concavity is a point in the spline model standing above the line connecting the previous and following point in the spline. A convexity is a point standing below such line. 

Consecutive concavities (convexities) define a concave (convex) section of the spline model. In Fig. \ref{fig:spline} the concave sections are highlighted with different filling colors.

\textbf{Peaks and shoulders: } Within a concave section of the spline, we identify one peak as the highest point internal to the concave section. If the highest point is at the boundary of the section, then the section contains no peaks. In this case, we identify a \textit{shoulder} as the point with the maximal residual from the line connecting the left and right boundary of the concave section. Such shoulder is then treated similarly to other peaks.

Each peak/shoulder is assigned a left and right uncertainty calculated using the position of the first points in the concave section that would bring down the residual of the peak below a given threshold ($5 \sigma$ in default options) if they were to replace the boundaries of the section. If no points satisfy this criterion, the left and right uncertainty are defined by the boundaries of the concave section.

\textbf{Prominence of the peaks: } The highest peak is assigned a prominence as the number of sigmas distinguishing it from the global minimum of the dataset. The other peaks are assigned a prominence with respect to the local minimum (if any) separating them from the highest peak. Shoulders are assigned a prominence with respect to the line connecting the boundaries of their concave section.

\textbf{Cross-matching peaks in different datasets: } The above procedure is performed separately on each available dataset. At this point we have a separate list of peaks for each dataset. The next task is to cross-match these lists.

We first identify as the same peak those peaks in different datasets whose positions fall within the uncertainty range of each other. Such duplicate peaks are replaced by a single peak with an uncertainty range defined as the intersection of the two original uncertainty ranges and the prominence taken as the sum of the prominences of the two peaks.

As a second step, we also identify as the same peak those peaks in which only one of the peak positions falls in the uncertainty range of the other peak. This is a weaker condition that is applied only after the first condition has already been used to refine all well-identified peaks.

Finally, all peaks with prominence below a chosen threshold ($10 \sigma$ in default options) are removed.

\textbf{Peak selection: } In the end, we save the positions $t_1$ and $t_2$ of the two peaks with the highest prominence as defined above. It is evident that one of these peaks may also be a shoulder to a main peak, provided they are separated by a convex section of the spline.

\textbf{Maximal asymmetry: } If this algorithm only finds one peak and no shoulders, for each dataset we calculate the deviation of each point in the spline models from the symmetric spline model obtained by reflecting the spline around the position of the lonely peak found. The maximal positive deviation is then taken as the second peak.

\vspace{0.2cm}

The procedure described here and adopted by \texttt{InitCond} is very effective for well-behaved datasets and allows to find features in the data that are above the prescribed threshold. Noise in the baseline may produce spurious peaks that can be removed by increasing the overall peak threshold. 

\subsection{Initial conditions for Single-lens-single-source models} \label{Sec init SLSS}

Single-lens-single-source microlensing events only depend on three minimal parameters: the time of closest approach $t_0$, the Einstein time $t_E$ fixing the timescale of the event, and the impact parameter $u_0$, determining the peak magnification. By default, we also fit for the source size $\rho_*$ in units of the Einstein radius $\theta_E$, which is however seldom constrained in microlensing events.

Once we have found the two most prominent peaks $t_1$ and $t_2$ from the previous algorithm, for single-lens-single-source events we just take $t_0=t_1$, i.e. we identify the position of the highest peak as the closest approach time between source and lens. The second peak/shoulder/asymmetry is ignored in these fits. All remaining parameters $u_0,t_E,\rho_*$ are taken from a grid since fitting of this class of models is so fast and the number of parameters is sufficiently small that this strategy remains the most efficient.


\subsection{Initial conditions for Single-lens-binary-source} \label{Sec init SLBS}

For single-lens-binary-source models, we have to duplicate the closest approach time and the impact parameters. We naturally take $t_{0,1}=t_1$, $t_{0,2}=t_2$, i.e. we match the positions of the two peaks identified in the previous step to the closest approach times between each source and the lens. One more parameter is the flux ratio between the two sources $q_f$. Furthermore, in principle, also the secondary source has its own finite radius, but it is convenient to calculate it from the radius of the primary source and rescaling it by some luminosity-radius relation in order to enforce consistency with the flux ratio. The choice $\rho_2 = \rho_1 q_f^{0.225}$ is just one among popular choices depending on the section of the main sequence we are considering. Whatever the relation used, it would be quite exceptional that both sources show finite-size effect at the same time. So, the particular choice of this relation is not expected to have any impact on the general model search performed by \texttt{RTModel}.

Similarly to single-source mdoels, the parameters $u_{0,1}, u_{0,2}$, $t_E, \rho_*, q_f$ are again taken from a grid. 

\subsection{Initial conditions for Binary-lens-single-source} \label{Sec init BLSS}

In this case, we have 3 more parameters with respect to the single-lens case: the mass ratio between the two lenses $q$, the separation $s$ in units of the Einstein angle, the angle $\alpha$ between the vector joining the two lenses (oriented toward the more massive one) and the source direction. We also notice that $t_0$ and $u_0$ are the closest approach parameters defined with respect to the center of mass of the lens. 

For binary lens models we use a template library currently made up of 113 different n-ples of parameters. Each template is characterized by its parameters $s,q,u_0,\alpha,\rho_*$. For each template we have pre-calculated the positions of two peaks in units of the Einstein time $t_{p1},t_{p2}$. The templates are chosen as representative light curves for regions in the parameter space in which the peaks have the same nature (same fold crossing or same cusp approach) \citep{Liebig}.  These templates including all caustic topologies, different mass ratios and different orientations of the source trajectory. Their parameters are listed in Tables \ref{Tab templates} and \ref{Tab templates2}. The idea is that the observed microlensing light curve should necessarily belong to one of these classes and thus the fit from a template in the same class should proceed without meeting any barriers in the chi square surface.

Therefore, for each template in the library, we set the initial conditions by copying the 5 parameters in the template, save for $t_E$ and $t_0$, which are determined by matching the peaks in the template $t_{p1},t_{p2}$ to the peak positions $t_1,t_2$ as found by the peak identification algorithm:
\begin{eqnarray}
&& t_E=\frac{t_2-t_1}{t_{p2}-t_{p1}}\\
&& t_0=t_{1}-t_E t_{p1}
\end{eqnarray}

We have two possible ways to match such times for each template (we may match $t_1$ to $t_{p1}$ and $t_2$ to $t_{p2}$ or exchange the two times). In one of the two cases we obtain a negative $t_E$, which is equivalent to inverting the source direction $\alpha$ and the sign of $u_0$. We finally end up with 226 different initial conditions for static binary lens models.


\begin{table}
\begin{tabular}{l|ccccccc}
\hline
 & $s$ & $q$ & $u_0$ & $\alpha$ & $\rho_*$ & $t_{p1}$ & $t_{p2}$ \\
 \hline 
\# 1 & 0.7 & 0.5 & 0.15 & 3.5 & 0.01 & -0.183 & 0.016 \\
\# 2 & 0.7 & 0.5 & 0.15 & 3.5 & 0.01 & -0.183 & 0.086 \\
\# 3 & 0.7 & 0.5 & 0.15 & 3.5 & 0.01 & 0.016 & 0.086 \\
\# 4 & 0.7 & 0.1 & 0. & 5.38 & 0.01 & -0.066 & 0.025 \\
\# 5 & 0.7 & 0.1 & 0. & 5.38 & 0.01 & -0.066 & 0.305 \\
\# 6 & 0.7 & 0.1 & 0. & 5.38 & 0.01 & 0.025 & 0.305 \\
\# 7 & 0.7 & 0.5 & 0. & 2. & 0.01 & -1.202 & -0.113 \\
\# 8 & 0.7 & 0.5 & 0. & 2. & 0.01 & -1.202 & 0.138 \\
\# 9 & 0.7 & 0.5 & 0. & 2. & 0.01 & -0.113 & 0.138 \\
\# 10 & 0.7 & 0.5 & -0.1 & 1.57 & 0.01 & -0.508 & -0.013 \\
\# 11 & 0.7 & 0.5 & -0.1 & 1.57 & 0.01 & -0.508 & 0.013 \\
\# 12 & 0.7 & 0.5 & -0.1 & 1.57 & 0.01 & -0.013 & 0.013 \\
\# 13 & 0.7 & 0.5 & 0.1 & 1.35 & 0.01 & -0.082 & 0.075 \\
\# 14 & 0.7 & 0.5 & 0.1 & 1.35 & 0.01 & -0.082 & 1.121 \\
\# 15 & 0.7 & 0.5 & 0.1 & 1.35 & 0.01 & 0.075 & 1.121 \\
\# 16 & 0.65 & 0.1 & 0.52 & 5. & 0.01 & -0.291 & 0.871 \\
\# 17 & 0.65 & 0.1 & 0.52 & 5. & 0.01 & -0.291 & 0.879 \\
\# 18 & 0.65 & 0.1 & 0.52 & 5. & 0.01 & 0.871 & 0.879 \\
\# 19 & 0.7 & 0.5 & -0.35 & 1.5 & 0.01 & -1.017 & -0.975 \\
\# 20 & 0.7 & 0.5 & -0.35 & 1.5 & 0.01 & -1.017 & 0.015 \\
\# 21 & 0.7 & 0.5 & -0.35 & 1.5 & 0.01 & -0.975 & 0.015 \\
\# 22 & 0.7 & 0.5 & 0.8 & 4. & 0.01 & 0.045 & 0.552 \\
\# 23 & 0.7 & 0.5 & 0.8 & 4. & 0.01 & 0.045 & 0.605 \\
\# 24 & 0.7 & 0.5 & 0.8 & 4. & 0.01 & 0.552 & 0.605 \\
\# 25 & 0.7 & 0.1 & 0.1 & 4.6 & 0.01 & -0.208 & -0.007 \\
\# 26 & 0.7 & 0.1 & 0.1 & 4.6 & 0.01 & -0.208 & 0.149 \\
\# 27 & 0.7 & 0.1 & 0.1 & 4.6 & 0.01 & -0.007 & 0.149 \\
\# 28 & 0.7 & 0.5 & -0.2 & 1.4 & 0.01 & -0.516 & 0.031 \\
\# 29 & 0.7 & 0.5 & -0.2 & 1.4 & 0.01 & -0.516 & 1.129 \\
\# 30 & 0.7 & 0.5 & -0.2 & 1.4 & 0.01 & 0.031 & 1.129 \\
\# 31 & 0.7 & 0.5 & 0.3 & 1.93 & 0.01 & -1.119 & 0.106 \\
\# 32 & 0.7 & 0.1 & 0.25 & 5.3 & 0.01 & -0.102 & 0.914 \\
\# 33 & 0.7 & 0.5 & 1.12 & 3.2 & 0.01 & -0.326 & -0.095 \\
\# 34 & 0.7 & 0.1 & 0.75 & 4.6 & 0.01 & -0.743 & 0.032 \\
\# 35 & 0.7 & 0.1 & 0.75 & 4.6 & 0.01 & -0.743 & 0.54 \\
\# 36 & 0.7 & 0.1 & 0.75 & 4.6 & 0.01 & 0.032 & 0.54 \\
\# 37 & 0.9 & 0.03 & -0.15 & 2. & 0.01 & -0.277 & 0.066 \\
\# 38 & 0.9 & 0.03 & -0.15 & 2. & 0.01 & -0.277 & 0.09 \\
\# 39 & 0.9 & 0.03 & -0.15 & 2. & 0.01 & 0.066 & 0.09 \\
\# 40 & 0.9 & 0.03 & -0.1 & 1.6 & 0.01 & -0.151 & -0.122 \\
\# 41 & 0.9 & 0.03 & -0.1 & 1.6 & 0.01 & -0.151 & 0.112 \\
\# 42 & 0.9 & 0.03 & -0.1 & 1.6 & 0.01 & -0.122 & 0.112 \\
\# 43 & 0.9 & 0.03 & -0.2 & 2.3 & 0.01 & -0.472 & 0.01 \\
\# 44 & 0.9 & 0.03 & -0.2 & 2.3 & 0.01 & -0.472 & 0.04 \\
\# 45 & 0.9 & 0.03 & -0.2 & 2.3 & 0.01 & 0.01 & 0.04 \\
\# 46 & 1.5 & 0.5 & 0.1 & 3.2 & 0.01 & -0.459 & -0.325 \\
\# 47 & 1.5 & 0.5 & 0.1 & 3.2 & 0.01 & -0.459 & 0.632 \\
\# 48 & 1.5 & 0.5 & 0.1 & 3.2 & 0.01 & -0.325 & 0.632 \\
\# 49 & 1.5 & 0.5 & 0. & 3. & 0.01 & -0.446 & -0.348 \\
\# 50 & 1.5 & 0.5 & 0. & 3. & 0.01 & -0.446 & 0.656 \\
\# 51 & 1.5 & 0.5 & 0. & 3. & 0.01 & -0.348 & 0.656 \\
\# 52 & 0.9 & 0.03 & -0.26 & 1.65 & 0.01 & -0.222 & 0.167 \\
\# 53 & 0.9 & 0.03 & -0.26 & 1.65 & 0.01 & -0.222 & 0.194 \\
\# 54 & 0.9 & 0.03 & -0.26 & 1.65 & 0.01 & 0.167 & 0.194 \\
\# 55 & 1.5 & 0.5 & 0.3 & 2.8 & 0.01 & 0.314 & 0.782 \\
\# 56 & 1.5 & 0.5 & 0.3 & 2.8 & 0.01 & 0.588 & 0.782 \\
\# 57 & 0.9 & 0.03 & -0.3 & 1.8 & 0.01 & -0.285 & 0.129 \\
\# 58 & 0.9 & 0.03 & -0.3 & 1.8 & 0.01 & -0.285 & 0.142 \\
\# 59 & 0.9 & 0.03 & -0.3 & 1.8 & 0.01 & 0.129 & 0.142 \\
\# 60 & 1.5 & 0.5 & 0.3 & 2.9 & 0.01 & -0.491 & 0.392 \\
\hline
\end{tabular}
\caption{Parameters of the templates used to set initial conditions (to be continued in Table 2).}
\label{Tab templates}
\end{table}

\begin{table}
\begin{tabular}{l|ccccccc}
\hline
\# 61 & 1.5 & 0.5 & 0.3 & 2.9 & 0.01 & -0.491 & 0.562 \\
\# 62 & 1.5 & 0.5 & 0.3 & 2.9 & 0.01 & 0.392 & 0.562 \\
\# 63 & 1.5 & 0.5 & 0.25 & 2.75 & 0.01 & 0.208 & 0.738 \\
\# 64 & 1.5 & 0.5 & 0.25 & 2.75 & 0.01 & 0.668 & 0.738 \\
\# 65 & 1.5 & 0.5 & 0.35 & 2.2 & 0.01 & 0.06 & 0.646 \\
\# 66 & 1.5 & 0.5 & 0.35 & 2.2 & 0.01 & 0.538 & 0.646 \\
\# 67 & 1.5 & 0.5 & 0.25 & 2.5 & 0.01 & -0.458 & 0.096 \\
\# 68 & 1.5 & 0.5 & 0.25 & 2.5 & 0.01 & -0.458 & 0.587 \\
\# 69 & 1.5 & 0.5 & 0.25 & 2.5 & 0.01 & 0.096 & 0.587 \\
\# 70 & 1.5 & 0.5 & 0.25 & 2. & 0.01 & -0.041 & 0.296 \\
\# 71 & 1.5 & 0.5 & 0.25 & 2. & 0.01 & -0.041 & 0.472 \\
\# 72 & 1.5 & 0.5 & 0.25 & 2. & 0.01 & 0.296 & 0.472 \\
\# 73 & 1.5 & 0.5 & 0.7 & 1.5 & 0.01 & -0.08 & -0.016 \\
\# 74 & 1.5 & 0.5 & 0.7 & 1.5 & 0.01 & -0.08 & 0.346 \\
\# 75 & 1.5 & 0.5 & 0.7 & 1.5 & 0.01 & -0.016 & 0.346 \\
\# 76 & 1.5 & 0.5 & 0.45 & 4.6 & 0.01 & -0.409 & -0.049 \\
\# 77 & 1.5 & 0.5 & 0.45 & 4.6 & 0.01 & -0.409 & 0.277 \\
\# 78 & 1.5 & 0.5 & 0.45 & 4.6 & 0.01 & -0.049 & 0.277 \\
\# 79 & 1.5 & 0.5 & 0.33 & 3.2 & 0.01 & -0.367 & 0.652 \\
\# 80 & 0.9 & 0.1 & -0.45 & 1.65 & 0.01 & -0.466 & -0.001 \\
\# 81 & 0.9 & 0.1 & -0.45 & 1.65 & 0.01 & -0.466 & 0.372 \\
\# 82 & 0.9 & 0.1 & -0.45 & 1.65 & 0.01 & -0.001 & 0.372 \\
\# 83 & 2.2 & 0.5 & 0.1 & 3. & 0.01 & 0.746 & 1.069 \\
\# 84 & 2.2 & 0.5 & 0.1 & 3. & 0.01 & 0.746 & 1.189 \\
\# 85 & 2.2 & 0.5 & 0.1 & 3. & 0.01 & 1.069 & 1.189 \\
\# 86 & 2.2 & 0.5 & 0.35 & 2.9 & 0.01 & 1.073 & 1.142 \\
\# 87 & 2.2 & 0.5 & 0.35 & 2.9 & 0.01 & 1.073 & 1.287 \\
\# 88 & 2.2 & 0.5 & 0.35 & 2.9 & 0.01 & 1.142 & 1.287 \\
\# 89 & 2.2 & 0.5 & 0.4 & 2.8 & 0.01 & 0.985 & 1.169 \\
\# 90 & 2.2 & 0.5 & 0.4 & 2.8 & 0.01 & 0.985 & 1.215 \\
\# 91 & 2.2 & 0.5 & 0.4 & 2.8 & 0.01 & 1.169 & 1.215 \\
\# 92 & 2.2 & 0.5 & 1.05 & 2. & 0.01 & 0.394 & 0.563 \\
\# 93 & 2.2 & 0.5 & 1.05 & 2. & 0.01 & 0.394 & 0.657 \\
\# 94 & 2.2 & 0.5 & 1.05 & 2. & 0.01 & 0.563 & 0.657 \\
\# 95 & 2.2 & 0.5 & 0.45 & 2.7 & 0.01 & -0.675 & 0.896 \\
\# 96 & 2.2 & 0.5 & 0.45 & 2.7 & 0.01 & -0.675 & 1.107 \\
\# 97 & 2.2 & 0.5 & 0.45 & 2.7 & 0.01 & 0.896 & 1.107 \\
\# 98 & 2.2 & 0.5 & 1. & 1.9 & 0.01 & 0.3 & 0.398 \\
\# 99 & 2.2 & 0.5 & 1. & 1.9 & 0.01 & 0.3 & 0.526 \\
\# 100 & 2.2 & 0.5 & 1. & 1.9 & 0.01 & 0.398 & 0.526 \\
\# 101 & 2.2 & 0.5 & 1.23 & 1.6 & 0.01 & -0.17 & 0.018 \\
\# 102 & 2.2 & 0.5 & 1.23 & 1.6 & 0.01 & -0.17 & 0.053 \\
\# 103 & 2.2 & 0.5 & 1.23 & 1.6 & 0.01 & 0.018 & 0.053 \\
\# 104 & 2.2 & 0.5 & 0.35 & 2.7 & 0.01 & -0.667 & 0.745 \\
\# 105 & 2.2 & 0.5 & 0.35 & 2.7 & 0.01 & -0.667 & 1.146 \\
\# 106 & 2.2 & 0.5 & 0.35 & 2.7 & 0.01 & 0.745 & 1.146 \\
\# 107 & 2.2 & 0.5 & 0.05 & 2.95 & 0.01 & -0.645 & 0.231 \\
\# 108 & 2.2 & 0.5 & 0.05 & 2.95 & 0.01 & -0.645 & 1.201 \\
\# 109 & 2.2 & 0.5 & 0.05 & 2.95 & 0.01 & 0.231 & 1.201 \\
\# 110 & 2.2 & 0.5 & 0.35 & 3. & 0.01 & -0.69 & 1.154 \\
\# 111 & 1.8 & 0.03 & 1. & 2.3 & 0.01 & -0.084 & 0.72 \\
\# 112 & 1.8 & 0.03 & 1. & 2.3 & 0.01 & -0.084 & 0.889 \\
\# 113 & 1.8 & 0.03 & 1. & 2.3 & 0.01 & 0.72 & 0.889 \\
\hline 
\end{tabular}
\caption{Parameters of the templates used to set initial conditions (continued from Table 1).}
\label{Tab templates2}
\end{table}

\begin{figure}
    \centering
    \includegraphics[width = 0.5\textwidth]{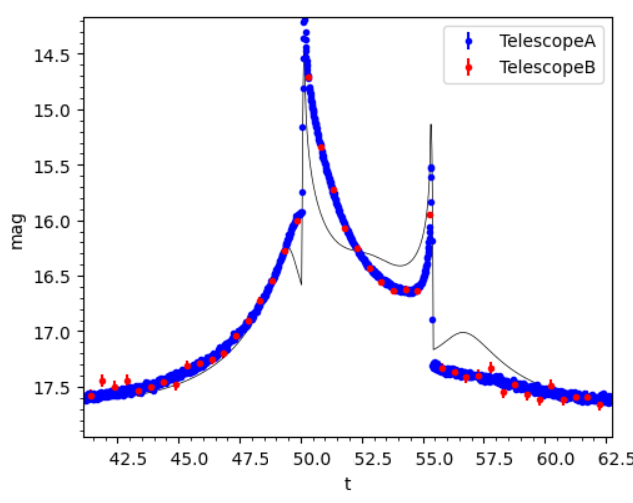}
    \caption{Example of template matching to some simulated data using template \# 47.}
    \label{fig:matching}
\end{figure}

Fig. \ref{fig:matching} illustrates how template matching works in initial conditions setting. Here we have some simulated data and one of the templates is matched to the data by using the main peaks, which in this case are caustic crossing peaks. With initial conditions set in this way, we are confident that at least one of the templates falls in the same light curve category of the observed event, so that the following fit will very quickly converge to the correct solution.

\subsection{Initial conditions for small planetary anomalies} \label{Sec init planetary}

The templates listed in tables \ref{Tab templates}-\ref{Tab templates2} cover all caustic topologies and source trajectories starting from mass ratios closer to 1, which are appropriate for stellar binaries. Although all planetary microlensing light curves are in principle reachable from these templates, the fitting algorithm needs to move a long way to reach such distant corners of the parameter space. As a boost for efficiency in the planetary regime, we also include further initial seeds determined along the lines initially proposed by \citet{GouldLoeb1992} (see also \citep{GaudiGould1997,Bozza1999,Han2006,Gaudi2012,Zhang2022}). This strategy starts from the parameters of the best models found from the single-lens-single-source search to inject the planetary perturbation at the position of the anomaly. It can only be implemented after the fits of the single-lens-single-source models, and then these further initial seeds are not calculated in the \texttt{InitCond} module, but in the \texttt{ModelSelector} (see \ref{Sec Model Selection}).

In practice, the parameters $t_0,t_E,u_0$ are taken from the single-lens-sigle-source model. The parameters $\rho_*$ is set to 0.001 because we want to be as sensitive as possible to small anomalies. The mass ratio is initially set to $q=0.001$, well in the planetary regime.

The other parameters are then obtained by requiring that the planet generates a peak in the position of the second peak $t_2$ as found by \texttt{InitCond}. In particular, we first calculate the abscissa of the anomaly along the trajectory
\begin{equation}
    \widehat{dt} = \frac{t_2-t_0}{t_E}.
\end{equation}

Then we find the angular distance of the anomaly from the primary lens 
\begin{equation}
    x_a = \sqrt{u_0^2 + \widehat{dt}^2}.
\end{equation}

Then we have the angle between the closest approach point and anomaly point
\begin{equation}
    \alpha_0 = \arctan \frac{u_0}{-\widehat{dt}}
\end{equation}

Now, let's distinguish the cases for close- and wide-separation planets.

\subsubsection{Wide-separation planets}
A wide-separation planet would produce a planetary caustic at distance $x_a$ if its separation is
\begin{equation}
    s_0 = \frac{1}{2} \left[ \sqrt{4+x_a^2} + x_a \right].
\end{equation}

We then decrease $q$ until the size of the planetary caustic (calculated as in \citep{Bozza2000,Han2006}) is smaller than the separation from the primary, namely, we require
$x_a < 4 \sqrt{q} / s_0^2$.

At this point, we just choose the separation of the planet in such a way that the planetary caustic is either just within or beyond the anomaly point:
\begin{equation}
    s_\pm = s_0 \pm \frac{4\sqrt{q}}{s_0^2},
\end{equation}
and take $\alpha =\alpha_0$ to fix all 7 parameters for the binary-lens model.

\begin{figure}
    \centering
    \includegraphics[width = 0.5\textwidth]{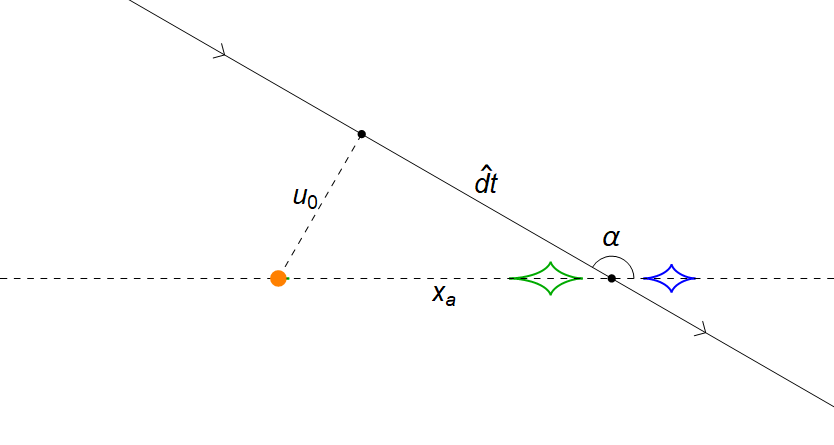}
    \caption{Determination of initial conditions for planetary fits in the wide configuration. The blue caustic comes with the choice $s=s_+$ and the green caustic with $s=s_-$.}
    \label{fig:planwide}
\end{figure}

The two values $s_\pm$ cover the inner/outer degeneracy typical of wide-separation planets \citep{Zhang2022}. Fig. \ref{fig:planwide} shows the two initial conditions obtained with this approach in a practical case, illustrating the geometric quantities introduces in this subsection.

\subsubsection{Close-separation planets}

For close-separation planets, a caustic in $x_a$ is obtained for
\begin{equation}
    s_0 = \frac{1}{2} \left[\sqrt{4+x_a^2} - x_a \right]
\end{equation}

Again we decrease $q$ until the size of the planetary caustic is less than the separation from the primary, namely, we require
$x_a < 2 * \sqrt(q) / s_0$ \citep{Bozza2000,Han2006}.

For close-in planets we fix the separation $s=s_0$, but we choose two values of $\alpha$ as
\begin{equation}
    \alpha_\pm = \alpha_0 + \pi \pm \arcsin \left| \frac{2 \sqrt{q}}{s x_a} \right |,
\end{equation}
which make the anomaly point in $t_2$ lie within or beyond one of the triangular caustics (note that $t_2$ is the time of a peak not a trough).

\begin{figure}
    \centering
    \includegraphics[width = 0.5\textwidth]{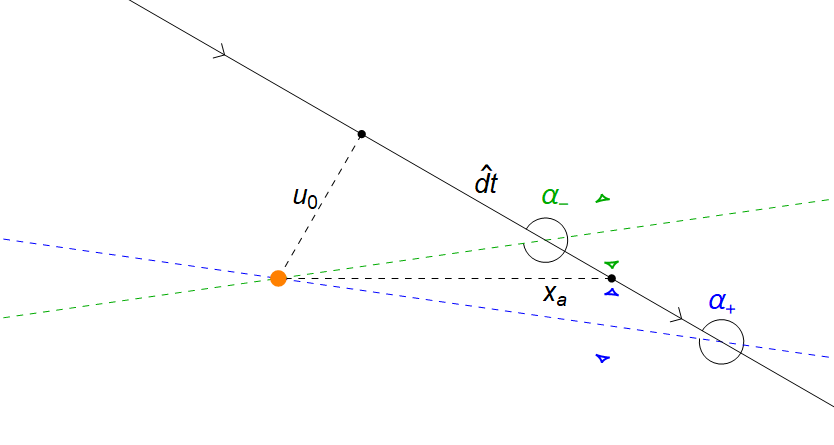}
    \caption{Determination of initial conditions for planetary fits in the close configuration. The blue caustic comes with the choice $\alpha=\alpha_+$ and the green caustic with $\alpha=\alpha_-$.}
    \label{fig:planclose}
\end{figure}

The two seeds obtained in this way are illustrated in Fig. \ref{fig:planclose}. Also in this case we cover the inner/outer degeneracy.

In total, for each single-lens-single-source model passing the \texttt{ModelSelector} selection, we have four additional initial conditions for the binary-lens fit well in the planetary regime.

\section{Fitting}

For each starting point as proposed by the initial conditions algorithm described in the previous section, we perform a downhill fit by Levenberg-Marquardt (LM) algorithm \citep{Levenberg,Marquardt} to find a local minimum of the chi square function. This is performed by launching the dedicated module \texttt{LevMar}, which fits a specific model (single-lens-single-source, single-lens-binary-source, binary-lens-single-source) from the chosen initial condition. Since we have a large number of initial conditions to test, it is convenient to launch many \texttt{LevMar} processes in parallel, exploiting the available cores. This is automatically managed by the main \texttt{RTModel} module, which launches a new process whenever one of the previous ones has terminated. 

The total computation time needed for modeling one microlensing event roughly scales with the inverse of the number of the available cores. It is however possible that one of the processes ends up in a region in which the computation becomes extremely slow (huge sources, extremely long Einstein times). For general purpose modeling, it is possible to set a time limit to avoid getting stuck because of these extreme regions of the parameter space. If the user believes that such regions need to be probed in depth, the time limit can be relaxed accordingly.

\subsection{The Levenberg-Marquardt algorithm}

The LM algorithm interpolates between steepest descent and Gauss-Newton's algorithm adaptively. The steepest descent just takes a step in the direction opposite to the gradient of the chi square function, whereas the Gauss-Newton's method attempts a parabolic fit of the chi square function and directly suggests the position of the local minimum. In principle, Gauss-Newton's method is a second-order method that is much more efficient than steepest descent. However, if we are still too far from the minimum, the parabolic approximation may be very poor and lead to weird suggestions. In the LM approach, a control parameter $\lambda$ modifies the set of equations that provide the step to the next point. When $\lambda$ is small, the equations behave like in Gauss-Newton, whereas the set becomes equivalent to steepest descent when $\lambda$ is large. The algorithm starts with an intermediate value of $\lambda$ and if the suggested point improves the chi square then $\lambda$ is lowered by a fixed factor. If the suggested point is worse than the previous one, $\lambda$ is increased until the suggested point improves the chi square.

The LM algorithm is very efficient in dealing with complicated chi square functions such those characterizing microlensing. In fact, it is able to follow long valleys in the chi square surface adapting the step size to the local curvature and then rapidly converge to the minimum when it gets close enough. Of course each step needs the evaluation of the local gradient, which requires numerical derivatives for each non-linear parameter. Therefore, the total number of microlensing light curves to be calculated is $(1+p+a)N_{steps}$, where $p$ is the number of non-linear parameters, $N_{steps}$ is the number of steps to achieve the convergence threshold and $a$ is the average number of adjustments to the $\lambda$ parameter per step, which is of order one. 

As stopping conditions for our fits we set a default (but customizable) maximum number of 50 steps or three consecutive steps with a relative improvement in $\chi^2$ lower than $10^{-3}\chi^2$.  In general, the latter condition is met in most fits well before the 50 steps, which only affect fits starting very far from minima. For each minimum found, we also report the local covariance ellipsoid calculated through inversion of the Fisher matrix. This is also used in model selection to identify and remove duplicates (see Section \ref{Sec Model Selection}).

\subsection{Microlensing computations}

All models are calculated by the public package \texttt{VBBinaryLensing} \citep{Bozza1,Bozza2,Bozza3}, which is now a separate popular spin-off of the original \texttt{RTModel} project\footnote{\url{https://github.com/valboz/VBBinaryLensing}}. Indeed, most of the credits for the efficiency of \texttt{RTModel} come from the fast and robust calculations provided by this package. On the other hand, in more than ten years of service, \texttt{RTModel} has ensured that the \texttt{VBBinaryLensing} package has been tested on an extremely large variety of microlensing events, enhancing its reliability. 

In short, \texttt{VBBinaryLensing} uses pre-calculated tables for single-lens calculations with finite source effect and the contour integration algorithm \citep{Dominik1995,Dominik1998,GouldGaucherel} for binary lenses. The idea is to sample the source boundary and invert the binary lens equation in order to obtain a sampling of the images boundaries. The inversion of the lens equation is performed by the Skowron \& Gould algorithm \citep{SkowronGould}. The sampling of the source boundary is driven by accurate error estimators that select those sections that need denser sampling. The accuracy is greatly improved by parabolic corrections, while limb darkening is taken into account by repeating the calculation on concentric annuli. The number and location of the annuli is also chosen dynamically on the basis of error estimators \citep{Bozza1}. Since finite-size calculations are computationally expensive, the code starts with a point-source evaluation and estimates the relevance of finite-source corrections by the quadrupole correction and tests on the ghost images \citep{Bozza2}. More recently, also the computation of the astrometric centroid has been added to  \texttt{VBBinaryLensing} \citep{Bozza3}, and will be incorporated in future releases of  \texttt{RTModel} for simultaneous fitting of photometric and astrometric microlensing.

 \texttt{RTModel} also inherits all specific techniques for higher order effect computations from \texttt{VBBinaryLensing}, including parallax, binary-lens orbital motion or binary-source xallarap. In particular, for space-based datasets, which have been marked by a satellite number in the data pre-processing phase (see Section \ref{Sec data}), the source is displaced according to the microlensing parallax vector components using knowledge of the satellite position at the time of the observation \citep{Gould1992ApJ...392..442G}. This is derived from ephemerides tables that can be easily downloaded by the user from the NASA Horizon website\footnote{\url{https://ssd.jpl.nasa.gov/horizons}}.

\subsection{Exploring multiple minima} \label{Sec Multiple Minima}

As stated before, the optimization problem in microlensing is complicated by the high dimensionality of the parameter space and the existence of caustics, which generate complicated barriers and valleys in the $\chi^2$ surface. The library of initial conditions presented in Section \ref{Sec InitCond} ensures that at least one initial condition should be in the correct region of the parameter space, with the same sequence of peaks and minima as the true solution. However, this is not always enough to guarantee that the correct initial condition will converge to the global minimum. In fact, accidental degeneracies due to the sampling in the data may fragment the correct region of the parameter space into multiple minima. Furthermore, the path to the global minimum may pass through extremely thin valleys surrounded by high barriers. Such valleys typically slow down the LM algorithm, which will take too many steps to converge, eventually stopping before the true minimum has been reached.

Optimization for not globally convex function has a very long history that we shall not repeat here. The strategy that we adopt to broaden our search and avoid stopping at the first local minimum is a Tabu-search type \citep{Glover1989,Glover1990}, in which past visited solutions are avoided. 

Whenever a local minimum $\boldsymbol{x}_0$ is found by the LM algorithm, a ``bumper'' is placed in its position that will repel new fits from the minimum found. The bumper is first defined to coincide with the covariance ellipsoid centered on the local minimum,  calculated through the inversion of the Fisher matrix. As the LM is repeated from the same initial condition, it will end up at some point $\boldsymbol{x}$ in the bumper region. At this point, the bumper will push the fit outside its ellipsoid region according to the following rule
\begin{equation}
    \hat{ \boldsymbol{x}} = \boldsymbol{x} -2p \frac{\boldsymbol{\Delta x}}{\sqrt{\boldsymbol{\Delta x}^T \hat A \boldsymbol{\Delta x} }}, \label{bumper}
\end{equation}
where $\boldsymbol{\Delta x}= \boldsymbol{x}-\boldsymbol{x}_0$ is the vector from the bumper center to the current position of the fit $\boldsymbol{x}$; $\hat A$ is the curvature matrix in $\boldsymbol{x}_0$, i.e. the inverse of the covariance matrix; $p$ is a coefficient that can be chosen by the user (initially set to 2).

With this push, the fit is sent outside the covariance ellipsoid of the minimum, but in general this is not sufficient to exit the attractor basin of the minimum. The bumper size is then increased by a fixed factor dividing the curvature $\hat A$, so that if the fit comes back close to the same minimum, it receives another push that brings it further way. Eventually, the fit will exit the attractor basin and converge to some other minimum (if any).

\begin{figure}
    \centering
    \includegraphics[width = 0.4\textwidth]{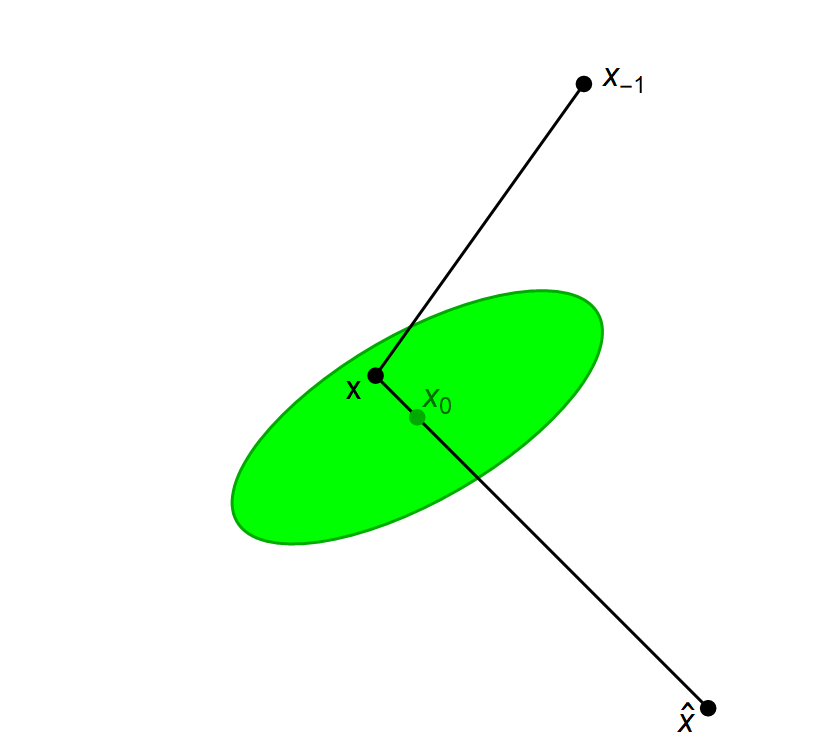}
    \caption{Bumper mechanism to jump out of local minima. The fit coming from $\mathbf{x}_{-1}$ ends in $\mathbf{x}$, within the covariance ellipsoid of the previously found minimum $\mathbf{x}_0$. It is then bumped to $\hat{\mathbf{x}}$, where it may continue the exploration to other minima.}
    \label{fig:bumper}
\end{figure}

Fig. \ref{fig:bumper} illustrates the bumper mechanism. We note that the minus sign in Eq. (\ref{bumper}) pushes the fit across the minimum to the other side. In principle, we could have made the opposite choice and bump the fit away from the minimum along the $\mathbf{\Delta x}$ vector, or even bump to a random direction. Our specific choice comes from the fact that a common situation occurring in fitting is that the fit is following some long valley but gets stuck in a local minimum. Yet the true minimum is just beyond this local minimum. Therefore, in the next attempt we want to pass beyond this minimum rather than being bounced back. Indeed, in our experience, we find that this choice is sensibly more productive in terms of new minima found.

\begin{figure}
    \centering
    \includegraphics[width = 0.5\textwidth]{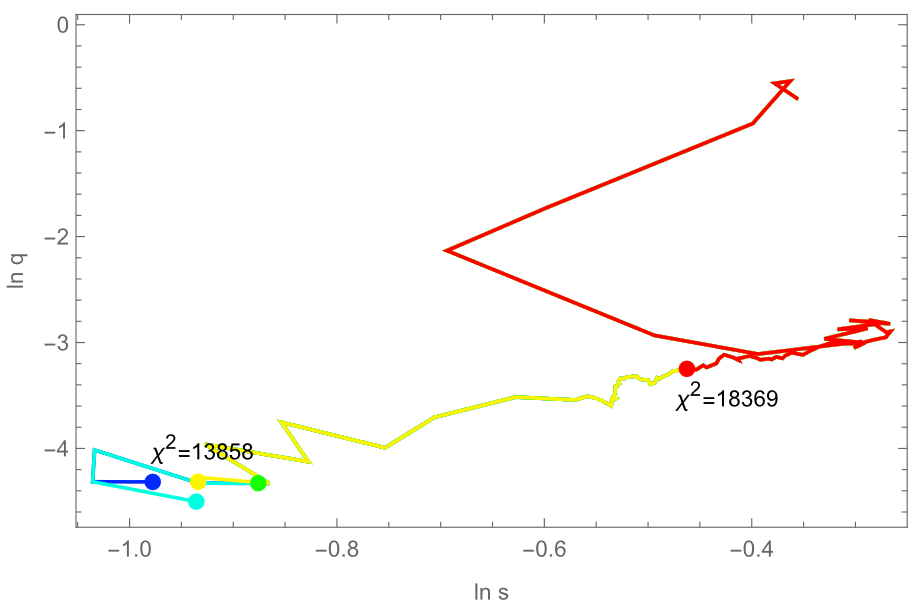}
    \caption{Fit track obtained on some simulated data projected to the plane $(\ln s, \ln q)$ in a binary-lens fit. We see that the first fit (red track) ended at a local minimum with high $\chi^2$. Then, repeating the fit with the bumper mechanism led to a different region with much lower $\chi^2$ where some more minima were found.}
    \label{fig:chain}
\end{figure}

Fig. \ref{fig:chain} illustrates the benefit of the bumper mechanism by showing fitting tracks starting from the same initial seed. The first fit (red track) gets stuck in a local minimum, but the second fit jumps out of the minimum thanks to the bumper mechanism and more minima are found with much lower chi square. The number of minima to be found from the same initial seed can be chosen by the user as usual (5 is the default choice).

\section{Model selection} \label{Sec Model Selection}


With so many initial conditions and the possibility to search for multiple minima from the same initial condition, as explained in Section \ref{Sec Multiple Minima}, \texttt{RTModel} accumulates a large number of models for each class. For example, for static binary lenses, we have 226 initial conditions producing many minima each (5 with the default settings explained above).

The module \texttt{ModelSelector} is designed to vet these models, remove duplicates, check for all possible reflections and retain the most interesting models. 

First of all, we consider two models with similar parameters as duplicates if their uncertainty ellipsoids (obtained by the covariance matrix) overlap. This criterion is very simple and allows to remove minima found by two different fits. However, it is still possible that models lying along a valley with high enough curvature in the parameter space survive this cut. At the level of preliminary model search, it is not a bad idea to retain such closely related models for following detailed local searches using MCMC. In fact, starting a Markov chain from two different points in the same valley is a recommended sanity check that all chains converge to the same model. As a control parameter, the user may adjust the threshold in the number of sigmas to consider two models as duplicates as desired.

Of course, the removal of duplicates takes into account all possible reflection symmetries in the class of models examined. For example, for static binary lenses, there is a reflection symmetry around the lens axis and we may also reflect the two lenses and the source trajectory around the center of mass. So, two models that only differ by a reflection symmetry are also considered as duplicates.

After the removal of duplicates, we have a list of (independent) models sorted by increasing $\chi^2$. Along with the best model, we consider as successful models all those that satisfy the condition
$\chi^2<\chi^2_{thr}$, with 
\begin{equation}
\chi^2_{thr}=\chi^2_{best}+ n_\sigma \sqrt{2 d.o.f.}. \label{chi2thr}
\end{equation}
With $n_\sigma=1$ this corresponds to one standard deviation in the $\chi^2$ distribution. The user may decide to be less or more generous by adjusting the parameter $n_\sigma$ here. 

Finally, the user may also specify a maximum number of models to be saved if there are too many solutions, e.g. for an ongoing event that has not developed any distinctive features yet. The selected models are then proposed as possible solutions for the given class of models analyzed by \texttt{ModelSelector}.

\subsection{Sequence of operations in model search} \label{Sec sequence}

As we have seen up to now, \texttt{LevMar} is a general fitter based on LM algorithm, using the bumper mechanism to jump out of local minima and continue the search towards other minima. It works with all light curve functions offered by \texttt{VBBinaryLensing}, including single- and binary-lens, single- and binary-source, parallax, xallarap or orbital motion.

For single-lens-single source models, the fit is extremely fast, so we just run \texttt{LevMar} from each initial condition in the grids discussed in Section \ref{Sec init SLSS}. Then, as \texttt{ModelSelector} selects the best models, it also creates initial conditions for single-lens-single-source models including parallax. In its default configuration, \texttt{RTModel} assumes that parallax is a small perturbation to static models. Therefore, it just starts from the best static models with zero parallax and lets the parallax components free to vary in \texttt{LevMar}. In the end, \texttt{ModelSelector} is run again to select the best models including parallax.

For single-lens-binary-source models the scheme is very similar. We run \texttt{LevMar} on all initial conditions from the grid described in Section \ref{Sec init SLBS}. Then \texttt{ModelSelector} selects the best models and sets the initial conditions for models of binary-sources including xallarap. In the end, \texttt{ModelSelector} is run again to select the best models in this category as well.

For binary-lens models, we run \texttt{LevMar} on all initial conditions from the template library described in Section \ref{Sec init BLSS}. We also add the planetary initial conditions described in \ref{Sec init planetary}, which are built when \texttt{ModelSelector} is run on single-lens-single-source models. After all fits are completed, we run \texttt{ModelSelector} to generate a list of viable independent models. The selected static binary-lens models are then taken as initial conditions for models including parallax and for models including parallax and orbital motion. Since parallax breaks the reflection symmetry around the lens axis, both mirror models are taken as independent initial conditions.

For models including parallax and models with parallax and orbital motion we remove duplicates and make the same selection process as before using \texttt{ModelSelector}.

In the end, we have a selection of models for each category, both single-lens and binary-lens, with different levels of higher orders. These models are ready for the final discussion and classification of the event (Sec. \ref{Sec classification}).

We note that the procedure just described assumes that parallax and orbital motion only make a small perturbation to a model that is already within the best choice for static binary-lens models. This is not always true. When the parallax is large or the orbital motion fast, the microlensing light curve can be dramatically modified, showing features that are impossible to reproduce with static binary models. In these cases, the whole approach of the template library built on rectilinear source trajectories fatally fails. 

There are two possible approaches to deal with such cases. The first is to skip the static binary-lens fitting and fit directly for models with parallax (and possibly orbital motion). This gives more freedom to the fits from the very beginning rather than checking for parallax on the best models only. The disadvantage is that if the parallax is not sufficiently constrained we may end up with very large unrealistic values. Therefore, we do not recommend taking this approach for all microlensing events, but only for those for which the perturbative approach fails.

The second is to fit only a section of the light curve that poses no problems to the perturbative approach and then add the remaining points gradually so that models including parallax and orbital motion are ``adiabatically'' adjusted to the new data. Indeed, fitting models in real-time has the advantage of catching good model(s) on the first part of the light curve, where higher orders can be neglected, and then follow the evolution of models in the parameter space as long as more data points are taken.

\section{Classification of the event} \label{Sec classification}


\begin{figure}
    \centering
    \includegraphics[width = 0.5\textwidth]{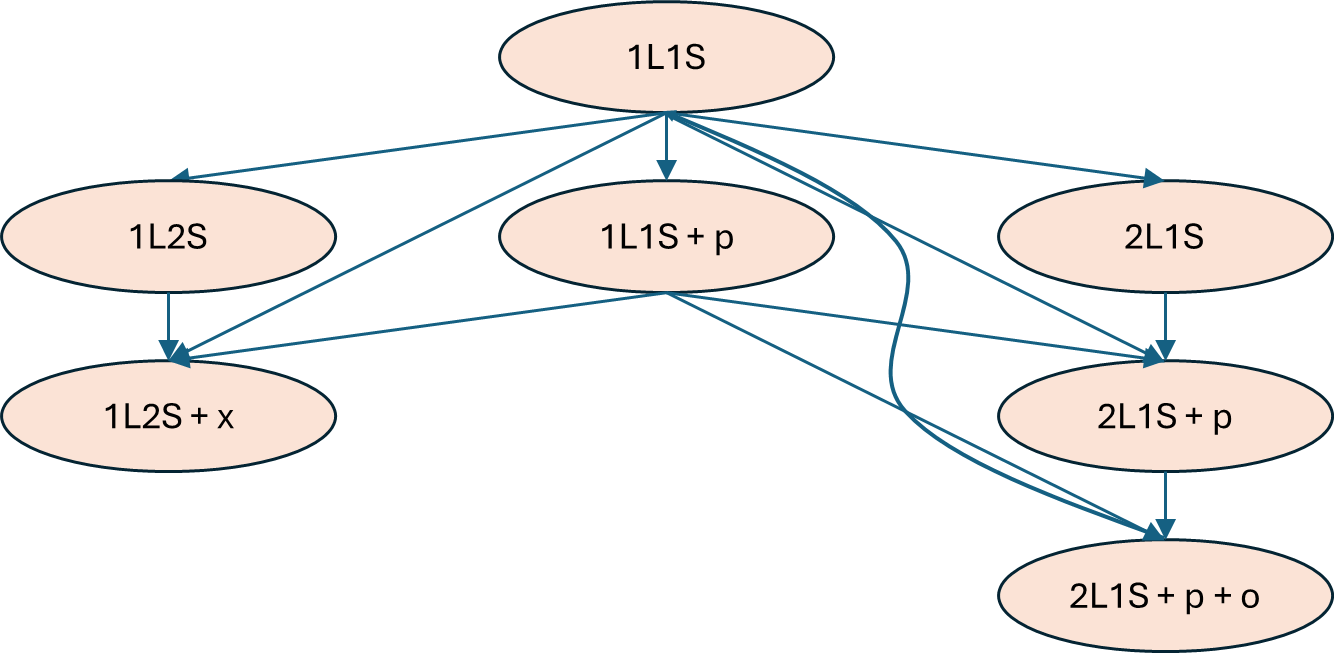}
    \caption{Hierarchy of model categories examined by \texttt{RTModel}. 1L1S means single-lens-single-source, 1L2S is single-lens-binary-source, 2L1S is binary-lens-single-source. "+p" indicates the presence of parallax, "+x" is xallarap, "+o" is lens orbital motion. The arrows go from lower-dimensionality models to higher-dimensionality models that include them as special cases (nested models).}
    \label{fig:hierarchy}
\end{figure}

The module \texttt{Finalizer} attempts a classification of the event based on comparison of the $\chi^2$ obtained by fitting different models. In this comparison we re-scale all $\chi^2$s by the factor $f=d.o.f./\chi^2_{best}$. In practice, we assume that error bars should be re-scaled by $f^{-1/2}$ as suggested by our best model. In the following tests, we assume that such re-scaling has been done.

\textbf{Nested models: } Most of the models we fit are nested. For example, the Single-Lens-Single-Source model is a special case of the binary lens in which one lens has vanishing mass. The static binary lens is a special case of the binary lens with parallax and orbital motion. Fig. \ref{fig:hierarchy} is a diagram showing the nesting relations among all model categories considered by \texttt{RTModel}. Every time we add parameters, we enlarge the parameter space increasing the freedom to adapt our model to the data. However, such improvement does not automatically imply evidence for the additional effect being necessary to explain the data. In fact, the additional parameters may just artificially adapt the model to the natural scatter of the data.

Following Wilks' theorem, the evidence for a model with $m$ additional parameters compared to a simpler model with less parameters (null hypothesis) is tracked by the $\chi^2$ distribution for $m$ degrees of freedom. We consider as a clear evidence a $\Delta\chi^2$ beyond the 0.999999426 threshold (corresponding to $5\sigma$) of the $\chi^2$ distribution. 

\begin{table}
\centering
\begin{tabular}{cc}
\hline
$m$ & $\Delta\chi^2$ \\
 \hline 
1 & 36. \\
2 & 40.0872 \\
3 & 43.4518 \\
4 & 46.4625 \\
5 & 49.2497 \\
6 & 51.878 \\
7 & 54.3854 \\
8 & 56.7964 \\
9 & 59.1282 \\
\hline 
\end{tabular}
\caption{$\chi^2$ thresholds at $5\sigma$ required to validate a test hypothesis with $m$ additional parameters with respect to a null hypothesis.}
\label{Tab:Wilks}
\end{table}

Models that do not pass this test compared to models with less parameters are discarded. Conversely, models with less parameters are dropped if there is a model with a higher number of parameters that passes this test. Table \ref{Tab:Wilks} lists the threshold indicated by Wilks theorem for small numbers of additional parameters.

\textbf{Non-Nested models: } The previous criterion applies to binary lens models compared to single-lens models as well as binary source models compared to single-source models. However, binary lens and binary source models are not nested one in the other and cannot be compared using such test. For non-nested model, we coherently adopt the same criterion used in \texttt{ModelSelector} to compile the list of independent models of a given class. We only retain models satisfying Eq. (\ref{chi2thr}). For example, if the best model is given by a binary lens, but a binary source model falls below $\chi^2_{thr}$, we consider this binary source model a viable alternative at the level of preliminary model search.

\textbf{Reported models: } It is clear that within a chain of nested models only models at a particular level will be reported, because the Wilks' theorem test will either remove models with more parameters but insufficient improvement or models with less parameters if models with more parameters perform better than the thresholds presented above. However, there might still be models on two different branches with comparable chi square (e.g. binary-source vs binary-lens) that will be retained. In definitive, successful models that pass the test for nested models are reported altogether as possible alternatives on different branches. In particular, the final list will also include degenerate binary lens models or binary source alternatives.

The list of viable models reported by \texttt{Finalizer} contains all combinations of parameters that explain the photometric data of the microlensing event light curve. We let the user select those models that also satisfy any known physical constraints external to the light curve analysis such as limits on blending flux, source radius, parallax, proper motion, astrometry, orbital motion and so on. Models can also be discriminated on the basis of Bayesian analysis with appropriate Galactic priors, but this is a level that goes beyond the current tasks assigned to \texttt{RTModel}, which just focuses on the search of preliminary models explaining the observed light curves. 

\section{WFIRST data challenge}
\label{Sec DataChallenge}

\texttt{RTModel} has been in activity since 2013 with hundreds of microlensing events analyzed in real time and visible on a public repository, as anticipated in the Introduction. A similar number has been analyzed offline within particular projects. Specific modules of \texttt{RTModel} have naturally evolved over these 10 years to enhance the effectiveness on all classes of microlensing events. 

An ideal opportunity to assess the effectiveness of \texttt{RTModel} approach came in 2018 by the WFIRST data challenge\footnote{\url{https://roman.ipac.caltech.edu/sims/Exoplanet_Data_Challenges.html}}. The goal of this competition was to stimulate new ideas for the analysis of massive microlensing data as expected from the future WFIRST (now {\it Roman}) mission\footnote{\url{https://www.jpl.nasa.gov/missions/the-nancy-grace-roman-space-telescope}}. This mission will make 6 continuous surveys of several fields of the Galactic bulge with the duration of about two months each \citep{Penny2019} and a cadence of 15 minutes. About 30 thousand microlensing events should be detected, with more than 1000 planets to be discovered in a wide range of masses, from Jupiters down to Mars-mass planets. This survey will complement Kepler statistics \citep{Borucki2010} in the outer regions of planetary systems beyond the so-called snow line \cite{Burn2021,Gaudi2021}. 

The analysis of several thousand microlensing events requires a software that is fast enough to ingest such a copious data flow and effective enough to discriminate true planets from contaminants and provide at least reasonable preliminary models that provide a good basis for further investigation. In the WFIRST data challenge, 293 light curves were simulated with the predicted cadence, the expected scatter in the photometry and all other survey specifications. These light curves included single-lens microlensing events, binary-lens, planetary-lens microlensing and some known contaminants, such as cataclismic variables. The participant teams were asked to provide an assessment for each light curve and a model.

\texttt{RTModel} took part in the data challenge proposing the only existing completely automatic algorithm running without any human intervention from the data preparation to the final assessment for each light curve. The success rate of our automatic interpretation and classification at that time was very encouraging and comparable to other platforms involving some human intervention or vetting at some level.

In the five years after that challenge, \texttt{RTModel} has continued its evolution. The version we are describing here and that is being released to the public has some important differences compared to the 2019 data challenge version. Nevertheless, we can still repeat the challenge using the same 293 simulated light curves and assess the performance of the current version on this well-established independent benchmark.

\begin{table*}
\centering
\begin{tabular}{|l|c|c|ccc|}
\hline
Class & Total & Successes &  Close/Wide & Undetected & Different model  \\
 \hline
 Single & 74 & 73 & 0 & 1 & 0 \\
 Binary ($q>0.03$) & 78 & 58 & 5 & 3 & 12 \\
 Planetary ($q<0.03$) & 48 & 37 & 7 & 1 & 3 \\
\hline 
\end{tabular}
\caption{Successes and failures of \texttt{RTModel} on the 2018 WFIRST data challenge using the default settings.}
\label{Tab data challenge}
\end{table*}

\begin{figure}
    \centering
    \includegraphics[width = 0.3\textwidth]{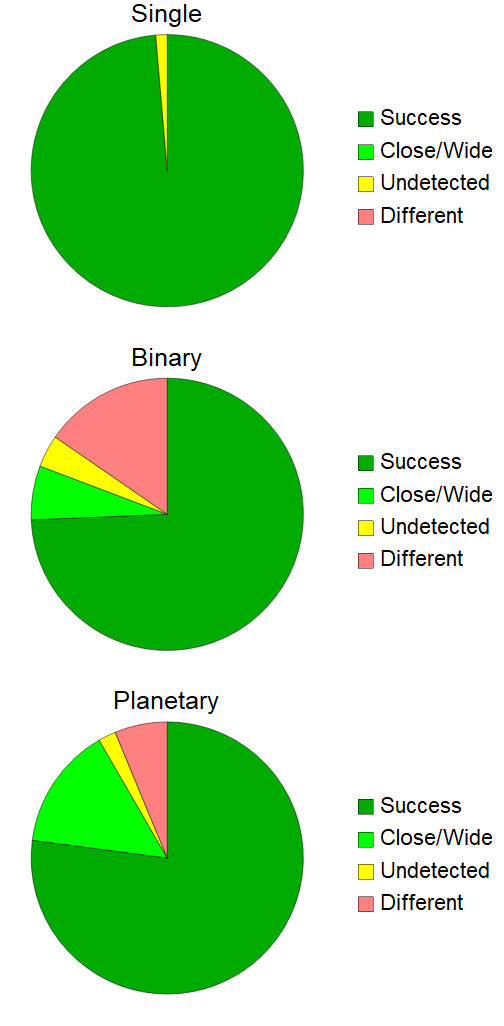}
    \caption{Pie charts for the results obtained by running the public version of \texttt{RTModel} with the default settings on the 2019 WFIRST data challenge (see Table \ref{Tab data challenge}).}
    \label{fig:pie}
\end{figure}

The success rates for single-lens, binary and planetary microlensing events are summarized in Table \ref{Tab data challenge} and visualized through pie charts in Fig. \ref{fig:pie}. The success rate is 98\% for single-lens events, 74\% for binary lenses with a mass ratio $q>0.03$ and 77\% for planetary lenses with $q<0.03$. These numbers might look relatively low for a modeling platform that promises efficient modeling for massive data flows, but let us first give a look at the ``failures'' before making our considerations. 

{\bf Close/wide: } In this column we have collected cases in which the data were simulated with a certain binary model and \texttt{RTModel} found a model with the same parameters except that the separation was the dual under the transformation $s \rightarrow 1/s$ \citep{Griest1998,Dominik1999,Bozza2000,An2005}. In principle, we would like to have both solutions in the final selection of proposed models, but \texttt{RTModel} did not find or discarded one of the two in its selection process. The impact of such loss is however marginal, since the planet is recovered anyway and detailed modeling after the preliminary search may also check the dual solution just to be sure that all possible models have been taken in consideration before the finalization of the analysis. So, depending on the analysis protocol, these events could be even included in the successful cases, raising the success rate to 80\% for binary and 91\% for planetary events.

{\bf Undetected: } Some anomalies in the simulated data are so subtle that were overlooked by \texttt{RTModel}. When the anomaly is at the noise level, a fully automatic platform may have a hard time in detecting and/or modeling such anomalies. Depending on the analysis protocol one adopts, such events may even pass undetected before they are sent to \texttt{RTModel}. So, if we remove them from our count, the success rate reaches 100\% for single-lens, 84\% for binary and 94\% for planetary events.

{\bf Different models:} These are the cases in which \texttt{RTModel} found a completely different model with respect to the simulated event. More investigation is needed to understand why the correct model was missed. In the case of planetary events, it might happen that the light curve can be also perfectly fit by a binary-lens model, as it is well known from several documented cases \citep{HanGaudi2008,Han2009}. For binary-lens events, we had some problems in recovering events with strong orbital motion. One reason may come from the fact that the simulated events were built with two-parameters orbital motion, which is notoriously unphysical \citep{Bozza3,Ma2022}. In such cases, these light curves should be rather removed from the challenge. On the other hand, we can reasonably expect that strong orbital motion cannot be recovered by the perturbative approach pursued by \texttt{RTModel} in its default configuration. For events in which higher order effects are too strong, \texttt{RTModel} should be set to include parallax and/or orbital motion from the very beginning of the search. 

\begin{table}
\centering
\begin{tabular}{lll}
\hline
Event & Class & Notes\\
 \hline
 9 & Binary & close instead of wide \\
18 & Binary & different binary model \\
20 & Binary & Good fit with single-lens with parallax \\
28 & Binary & different binary model \\
40 & Planetary & wide instead of close \\
58 & Binary & close instead of wide \\
66 & Planetary & close instead of wide \\
68 & Binary & different binary model \\
69 & Planetary & different planetary model \\
72 & Binary & different binary model \\
76 & Binary & different binary model \\
92 & Planetary & inner instead of outer \\
107 & Planetary & Very weak primary signal \\
111 & Binary & different binary model \\
131 & Planetary & too small anomaly \\
147 & Binary & too low signal \\
175 & Binary & different binary model \\
182 & Binary & close instead of wide \\
184 & Binary & different binary model \\
189 & Single & peak below noise level \\
190 & Binary & different binary model \\
207 & Binary & different binary model \\
208 & Planetary & close instead of wide \\
223 & Planetary & close instead of wide \\
226 & Planetary & different binary model \\
253 & Planetary & binary model found \\
259 & Binary & different binary model \\
267 & Planetary & inner instead of outer \\
282 & Binary & close instead of wide \\
283 & Binary & weak anomaly \\
287 & Binary & extreme orbital motion \\
291 & Binary & close instead of wide \\
\hline 
\end{tabular}
\caption{Events in the data challenge for which \texttt{RTModel} does not return the correct model with the default options. }
\label{Tab failures}
\end{table}

 All events for which the model was not perfectly found by \texttt{RTModel} are collected in Table \ref{Tab failures}. Apart from the well-known degeneracies discussed before, many events (in particular binaries) suffer from a different treatment of orbital motion. With a success rate ranging from 74\% to 94\%, depending on the metrics we prefer to adopt, we consider the results obtained with the default settings of \texttt{RTModel} extremely encouraging. By tuning the options in some specific way, most of the missed models can be promptly recovered with little more investment in computational time. Of course, there are still good margins for improvement for \texttt{RTModel} in some particular limits, as discussed above. This is one of the commitments that we take for future versions. 


\section{Conclusions}

One of the things that makes microlensing so exciting is in the difficulties of modeling. The way how a combination of caustics and particular source trajectories may conspire to create spectacular brightening and dimming episodes is somewhat magical. Furthermore, coding what human intuition sees in these patterns into a general software that is able to deal with all possible realistic cases remains a formidable challenge. Grid searches on the vast microlensing parameter space requires expensive clusters and may be unscalable to large data flows as those expected from the {\it Roman} mission. It is thus important to invest in alternative algorithms that drive the fit after a preliminary analysis of the features appearing in the data. 

This is exactly the philosophy of \texttt{RTModel}, deployed in 2013 to model microlensing events in real time with a fast Levenberg-Marquardt algorithm from a set of templates covering all possible categories of microlensing light curves. These templates are matched to features recognized in the light curve and provide efficient initial conditions for fitting. The bumper mechanism acting at minima previously found broadens the exploration of the parameter space. Finally, we should not forget that \texttt{VBBinaryLensing} was created within the original \texttt{RTModel} project and was then made public as a separate appreciated spin-off.

\texttt{RTModel} also proposes an automatic classification of the light curve based on statistical thresholds to be applied to the $\chi^2$ obtained with different model categories. This can be particularly useful when dealing with large number of light curves that cannot be visually inspected one by one.

As we have seen after the WFIRST data challenge, there are margins for improvement of \texttt{RTModel}, in particular for low-signal anomalies and for long binary events that cannot be recovered as perturbations of static models. Alternative parameterizations can also be useful in particular cases. Future developments may aim at incorporating new algorithms in the general architecture of the software recovering these more tricky events and approaching a 100\% efficiency. In this respect, more focused simulations may be helpful to single-out those situations where the general algorithm is less efficient.

The modular structure of \texttt{RTModel} makes it very flexible to further additions or replacements of individual steps in the modeling run. The same template library can be customized or extended by users to improve the efficiency on specific situations. Future extensions that will be reasonably achieved in the mid/long-term include astrometric microlensing, Markov chain exploration, Bayesian analysis with interface to Galactic models, triple and/or multiple lenses and/or binary sources.

In addition to these, the publication of our algorithms will allow future microlensing pipelines to benefit from the long experience gained by \texttt{RTModel} and design even more efficient platforms. Indeed, we believe that \texttt{RTModel} will stand as a reference platform in microlensing for general-purpose modeling for many years.

\begin{acknowledgements}
We thank Greg Olmschenck for a revision and refinement of the installation of the \texttt{RTModel} package to make it as cross-platform as possible.

We also thank Etienne Bachelet and Fran Bartolic for help and advice spreading from \texttt{VBBinaryLensing} to \texttt{RTModel}.

We acknowledge financial support from PRIN2022 CUP D53D23002590006.
\end{acknowledgements}


\bibliographystyle{aa} 
\bibliography{RTModel.bbl}

\end{document}